\documentclass[a4paper,aps,twocolumn]{revtex4}
\RequirePackage[english]{babel}
\RequirePackage[latin1]{inputenc}
\RequirePackage[T1]{fontenc}
\RequirePackage{mathrsfs}
\RequirePackage{amsmath}
\RequirePackage{amssymb}
\RequirePackage{amsbsy}
\RequirePackage{color}
\RequirePackage{bm}
\begin{document}
\title{\bf{Singularity-free spinors in gravity with propagating torsion}}
\author{Luca Fabbri}
\affiliation{INFN \& dipartimento di fisica, universit{\`a} di Bologna,\\
via Irnerio 46, 40126 Bologna, ITALY}
\date{\today}
\begin{abstract}
We consider the most general renormalizable theory of propagating torsion in Einstein gravity for the Dirac matter distribution and we demonstrate that in this case torsion is a massive axial-vector field whose coupling to the spinor gives rise to conditions in terms of which gravitational singularities are not bound to form; we discuss how our results improve those that are presented in the existing literature, and that no further improvement can be achieved unless one is ready to re-evaluate some considerations on the renormalizability of the theory.
\end{abstract}
\maketitle
\section{Introduction}
With aLIGO's recent detection of gravitational waves, we finally completed the list of original tests that Einstein gravity had to pass: with all predictions successfully confirmed and being built on properly defined mathematical foundations, Einstein gravitation is the only physical construction that could claim to be both phenomenologically very accurate and theoretically well established. The sole problem that Einstein gravity may have is the theoretical possibility that singularities can form at high energies as predicted by the well-known Penrose-Hawking theorem.

To have Einstein gravity purged of this issue, the most straightforward manner is to render the Penrose-Hawking theorem inapplicable by making its hypotheses fail, and because these hypotheses are given by some conditions on the energy density tensor, one has to make these energy conditions not verified by the matter distributions.

A first and most natural way in which this can be done is to allow the space-time to display a torsion \cite{C1,C2,C3,C4}: when beside curvature torsion is present, Einstein gravity can be completed to a theory where beside the usual energy-curvature field equations there are new torsion-spin field equations \cite{S,K}, called Einstein--Sciama-Kibble theory of gravitation. In this ESK gravity Einstein field equations are modified in their source by the fact that the energy density tensor now contains also the presence of the spin density tensor \cite{h-h-k-n}, and consequently the energy conditions undergo to a corresponding modification due to spin.

This idea was the original thrust for several treatments of the problem back in the 1970s \cite{kop,taf,tra,s-h}, but although all these attempts were seemingly promising, nonetheless all of them were using spinning dust, a matter distribution that in some context may be a good approximation of a fundamental matter distribution, but which is obviously not a fundamental matter distribution in itself.

Because at the high energy at which singularities are supposed to form quantum effects should become relevant then it is necessary to employ matter distributions given by spinorial fields; in addition, the ESK gravity is already equipped to contain matter distributions having energy as well as spin, and therefore to host spinors. In this case the energy density tensor is modified as claimed above.

Taking into account Dirac spinorial fields was done in the 1970s and first of all by Kerlick in \cite{ke}, but perhaps surprisingly, what Kerlick found is that even in presence of torsion-spin coupling, when spinors were considered, the energy density tensor enhanced the energy conditions and hence resulting into worsening the problem. Another attempt has been made some year later by employing a radically different method of evaluation \cite{Inomata:1976wi}, but still its conclusion is inevitably the same as that of Kerlick.

This issue comes from the fact that in the ESK theory, the Dirac Lagrangian is such that the spin-torsion force has a repulsive character: a repulsive force comes from a positive potential, which increases the energy content as source of gravity, enhancing gravitational singularity. 

The repulsiveness of the torsion-spin interaction is due to the fact that the torsion-spin coupling constant is the Newton constant, and this is due to the fact that passing from the Einstein action to the Einstein--Sciama-Kibble action we simply take the Ricci scalar $\mathscr{L}\!=\!R(g)$ and allow torsion to be included by considering the torsionful Ricci scalar $\mathscr{L}\!=\!G(g,Q)$ which, despite containing the torsion, still consists of a single term: one single term requires one constant, which is already fixed as the Newton constant, making the torsional force repulsive in a necessary way.

Of course, there is no reason to limit ourselves to such simple action, and in fact, because torsion is a tensor that is subject to no gauge transformation, we can add square torsion terms to the action as $\mathscr{L}\!=\!G(g,Q)\!+\!Q^{2}$ which, having two terms, can have two coupling constants \cite{Fabbri:2011kq}.

Because the spin-torsion coupling constant is not compelled to be the Newton constant, being a new constant, it could even have a different sign. In this circumstance, the torsional force would turn into an attraction and the singularity formation might finally be circumvented.

This is the argument from which recently a variety of papers has come out \cite{Magueijo:2012ug,Khriplovich:2013tqa,Alexander:2014eva,Buchbinder:1985ym}, all encompassing singularity avoidance and the possibility of Big Bounce cosmologies.

This is promising, but once more, there are issues: on the one hand, although in all these theories the torsional forces can be attractive, the fact that they actually are is still a choice; on the other hand, these theories are the most extended ESK gravity, but the ESK gravity itself is not the most general theory, because the torsion does not propagate and torsional forces are non-renormalizable.

A renormalizable propagating torsion theory is the one presented in \cite{Carroll:1994dq}, but in this reference the authors restrict their Lagrangian to a propagating torsion term that is the square of the divergence of torsion; as it can be shown by employing the Velo-Zwanziger method, one can demonstrate that for such a term problems may arise, stretching from mismatch between number of field equations and degrees of freedom, failure to propagate or acausality.

The sole consistent renormalizable propagating torsion term is the square of the curl of torsion \cite{Fabbri:2014dxa}, and for such a theory the question we must now ask is whether or not the torsion-spin interaction is attractive necessarily.

In this paper we will study such a situation.
\section{Propagating torsion for gravitating spinors}
In this paper, we will refer to \cite{Fabbri:2014dxa} for all notations and in particular we recall that the torsion may be assumed to be completely antisymmetric without losing any generality whatsoever: then a completely antisymmetric torsion is equivalent to the dual of an axial vector $W^{\mu}$ whose curl will be designated with $(\partial W)_{\alpha\nu}$ as usual. As torsion is a tensor subject to no gauge transformation, we can always have it separated from the rest of the connection in every covariant derivative and curvature, and thus it is possible to write each covariant derivative and curvature in their torsionless form accounting for torsion as an additional tensor field. The $R_{\mu\nu}$ is the torsionless Ricci curvature.

The $\boldsymbol{\gamma}^{\mu}$ are Clifford matrices with $\boldsymbol{\pi}\!=\!i\boldsymbol{\gamma}^{0} \boldsymbol{\gamma}^{1} \boldsymbol{\gamma}^{2} \boldsymbol{\gamma}^{3}$ being the parity-odd matrix. The $\overline{\psi}$ and $\psi$ are Dirac conjugate spinors. And $\boldsymbol{\nabla}_{\mu}$ is the torsionless covariant derivative.

In reference \cite{Fabbri:2014dxa} it has also been proven, by following the Velo-Zwanziger method, that the most general kinetic term of torsion must have the structure of the torsion curl square invariant. Given this, an easy inventory of all the other terms leads to the total Lagrangian given by
\begin{eqnarray}
\nonumber
&\mathscr{L}\!=\!\frac{1}{4}(\partial W)^{2}\!-\!\frac{1}{2}M^{2}W^{2}\!+\!R-\\
&-i\overline{\psi}\boldsymbol{\gamma}^{\mu}\boldsymbol{\nabla}_{\mu}\psi
\!+\!X\overline{\psi}\boldsymbol{\gamma}^{\mu}\boldsymbol{\pi}\psi W_{\mu}
\!+\!m\overline{\psi}\psi
\label{l}
\end{eqnarray}
as the Lagrangian we will employ. Having $3$ independent fields we could normalize $3$ coefficients, the $3$ remaining coefficients being given by $X$ as coupling between torsion and spin and $M$ and $m$ as torsion and spinor masses.

By varying this Lagrangian we obtain the torsion axial-vector field equations according to the expression
\begin{eqnarray}
&\nabla_{\rho}(\partial W)^{\rho\mu}\!+\!M^{2}W^{\mu}
\!=\!X\overline{\psi}\boldsymbol{\gamma}^{\mu}\boldsymbol{\pi}\psi
\label{torsionfieldequations}
\end{eqnarray}
with the gravitational field equations as
\begin{eqnarray}
\nonumber
&R^{\rho\sigma}\!-\!\frac{1}{2}Rg^{\rho\sigma}\!=\!\frac{1}{2}[\frac{1}{4}(\partial W)^{2}g^{\rho\sigma}
\!-\!(\partial W)^{\sigma\alpha}(\partial W)^{\rho}_{\phantom{\rho}\alpha}+\\
\nonumber
&+M^{2}(W^{\rho}W^{\sigma}\!-\!\frac{1}{2}W^{2}g^{\rho\sigma})+\\
\nonumber
&+\frac{i}{4}(\overline{\psi}\boldsymbol{\gamma}^{\rho}\boldsymbol{\nabla}^{\sigma}\psi
\!-\!\boldsymbol{\nabla}^{\sigma}\overline{\psi}\boldsymbol{\gamma}^{\rho}\psi
\!+\!\overline{\psi}\boldsymbol{\gamma}^{\sigma}\boldsymbol{\nabla}^{\rho}\psi
\!-\!\boldsymbol{\nabla}^{\rho}\overline{\psi}\boldsymbol{\gamma}^{\sigma}\psi)-\\
&-\frac{1}{2}X(W^{\sigma}\overline{\psi}\boldsymbol{\gamma}^{\rho}\boldsymbol{\pi}\psi
\!+\!W^{\rho}\overline{\psi}\boldsymbol{\gamma}^{\sigma}\boldsymbol{\pi}\psi)]
\label{metricfieldequations}
\end{eqnarray}
and finally the spinorial field equations are
\begin{eqnarray}
&i\boldsymbol{\gamma}^{\mu}\boldsymbol{\nabla}_{\mu}\psi
\!-\!XW_{\sigma}\boldsymbol{\gamma}^{\sigma}\boldsymbol{\pi}\psi\!-\!m\psi\!=\!0
\label{spinorfieldequation}
\end{eqnarray}
and that is the Dirac field equations: we do remark that the torsion axial-vector field equations have the structure of Proca equations, and thus the torsion tensor turns out to be equivalent to an axial-vector massive field, with the corresponding Proca energy density tensor contribution as source of gravitational field equations. For (\ref{l}) we have already discussed how many coefficients can and cannot be normalized, but one point to stress is that, regardless the normalization of the absolute value, it is not possible to fix the sign of these coefficients, and as for the signs we still have to fix $5$ of them: the sign of $X$ remains free as it is usual, and so we are down to $4$ signs; the relative sign between kinetic and mass terms is fixed by the positivity of the mass itself, and because this must happen not only for the spinor but also for the torsion, then we are down to $2$ signs left; the relative sign between the kinetic terms and the Ricci scalar is fixed by the positivity of the energy density tensor, and this has to occur not only for spinors but also for torsion, so we have no sign left to fix.

By taking the divergence of (\ref{torsionfieldequations}) and by contracting (\ref{metricfieldequations}), and employing (\ref{spinorfieldequation}), we get the following expressions
\begin{eqnarray}
&M^{2}\nabla_{\mu}W^{\mu}\!=\!2Xmi\overline{\psi}\boldsymbol{\pi}\psi
\end{eqnarray}
and
\begin{eqnarray}
&-2R\!=\!-M^{2}W^{2}\!+\!m\overline{\psi}\psi
\label{contraction}
\end{eqnarray}
the former as the partially conserved axial-vector current for torsion and the latter such that if substituted back into (\ref{metricfieldequations}) it allows to write the gravitational equations as
\begin{eqnarray}
\nonumber
&R^{\rho\sigma}\!=\!\frac{1}{2}[\frac{1}{4}(\partial W)^{2}g^{\rho\sigma}
\!-\!(\partial W)^{\sigma\alpha}(\partial W)^{\rho}_{\phantom{\rho}\alpha}+\\
\nonumber
&+\frac{i}{4}(\overline{\psi}\boldsymbol{\gamma}^{\rho}\boldsymbol{\nabla}^{\sigma}\psi
\!-\!\boldsymbol{\nabla}^{\sigma}\overline{\psi}\boldsymbol{\gamma}^{\rho}\psi
\!+\!\overline{\psi}\boldsymbol{\gamma}^{\sigma}\boldsymbol{\nabla}^{\rho}\psi
\!-\!\boldsymbol{\nabla}^{\rho}\overline{\psi}\boldsymbol{\gamma}^{\sigma}\psi)+\\
\nonumber
&+M^{2}W^{\rho}W^{\sigma}\!-\!\frac{1}{2}m\overline{\psi}\psi g^{\rho\sigma}-\\
&-\frac{1}{2}X(W^{\sigma}\overline{\psi}\boldsymbol{\gamma}^{\rho}\boldsymbol{\pi}\psi
\!+\!W^{\rho}\overline{\psi}\boldsymbol{\gamma}^{\sigma}\boldsymbol{\pi}\psi)]
\end{eqnarray}
which are equivalent to the original ones but best suited for the Penrose-Hawking singularity theorems.

In fact, for the Penrose-Hawking singularity theorems to engage, one has to make sure that some condition on the energy be verified: the strongest reads
\begin{eqnarray}
&R^{\rho\sigma}u_{\rho}u_{\sigma}\!\geqslant\!0
\end{eqnarray}
where $u^{\alpha}$ are time-like vectors, properly chosen. 

In our case, we have that
\begin{eqnarray}
\nonumber
&[\frac{1}{4}(\partial W)^{2}g^{\rho\sigma}
\!-\!(\partial W)^{\sigma\alpha}(\partial W)^{\rho}_{\phantom{\rho}\alpha}+\\
\nonumber
&+\frac{i}{2}(\overline{\psi}\boldsymbol{\gamma}^{\rho}\boldsymbol{\nabla}^{\sigma}\psi
\!-\!\boldsymbol{\nabla}^{\sigma}\overline{\psi}\boldsymbol{\gamma}^{\rho}\psi)+\\
\nonumber
&+M^{2}W^{\rho}W^{\sigma}\!-\!\frac{1}{2}m\overline{\psi}\psi g^{\rho\sigma}-\\
&-XW^{\sigma}\overline{\psi}\boldsymbol{\gamma}^{\rho}\boldsymbol{\pi}\psi]u_{\rho}u_{\sigma}
\!\geqslant\!0
\label{condition}
\end{eqnarray}
and because $u^{\alpha}$ is time-like it is possible to find a frame in which it only has the time component, where (\ref{condition}) becomes
\begin{eqnarray}
\nonumber
&\frac{1}{4}(\partial W)^{2}\!-\!(\partial W)^{0k}(\partial W)^{0}_{\phantom{0}k}+\\
\nonumber
&+\frac{i}{2}(\vec{\boldsymbol{\nabla}}\overline{\psi}\!\cdot\!\vec{\boldsymbol{\gamma}}\psi
\!-\!\overline{\psi}\vec{\boldsymbol{\gamma}}\!\cdot\!\vec{\boldsymbol{\nabla}}\psi)+\\
\nonumber
&+M^{2}|W^{0}|^{2}\!+\!\frac{1}{2}m\overline{\psi}\psi-\\
&-X\vec{W}\!\cdot\!\overline{\psi}\vec{\boldsymbol{\gamma}}\boldsymbol{\pi}\psi\!\geqslant\!0
\label{conditionspecial}
\end{eqnarray}
in which we clearly see the kinetic energies, mass terms and interactions of the torsion and the spinor field.

With this theory at our disposal we may now proceed in studying its consequences for the singularities.
\section{Avoiding singularities}
The first thing to check, as we have prepared the reader in the introduction, is whether or not in (\ref{l}) the resulting torsion-spin interaction is necessarily attractive.

In lack of exact solutions, our best chances are in seeing what happens in the effective approximation.

Because the torsion is massive, such an approximation can always be done, reducing (\ref{torsionfieldequations}) down to
\begin{eqnarray}
&M^{2}W^{\mu}\!\approx\!X\overline{\psi}\boldsymbol{\gamma}^{\mu}\boldsymbol{\pi}\psi
\label{a}
\end{eqnarray}
which allows us to integrate torsional degrees of freedom away; the energy condition (\ref{conditionspecial}) thus becomes
\begin{eqnarray}
\nonumber
&\frac{i}{2}(\vec{\boldsymbol{\nabla}}\overline{\psi}\!\cdot\!\vec{\boldsymbol{\gamma}}\psi
\!-\!\overline{\psi}\vec{\boldsymbol{\gamma}}\!\cdot\!\vec{\boldsymbol{\nabla}}\psi)
\!-\!\frac{X^{2}}{M^{2}}|i\overline{\psi}\boldsymbol{\pi}\psi|^{2}-\\
&-\frac{X^{2}}{M^{2}}|\overline{\psi}\psi|^{2}\!+\!\frac{1}{2}m\overline{\psi}\psi\!\geqslant\!0
\label{conditionveryspecial}
\end{eqnarray}
where spinors have been re-arranged with the Fierz identities given by $-\overline{\psi}\boldsymbol{\gamma}^{\mu}\boldsymbol{\pi}\psi \overline{\psi}\boldsymbol{\gamma}_{\mu}\boldsymbol{\pi}\psi\!=\!
|i\overline{\psi}\boldsymbol{\pi}\psi|^{2}\!+\!|\overline{\psi}\psi|^{2}$ therefore showing that the spin-torsion potential is negative and so the interaction is attractive necessarily.

It is important to notice that the attractiveness of the interaction comes directly from having fixed all the signs in the Lagrangian so to have the positivity of mass and energy of torsion, and thus ultimately from the fact that torsion propagates; without propagation of torsion, there can be no concept of energy or mass, then the sign of the torsional coupling constant could never be fixed, so that fixing it would require an arbitrary choice. None of this arbitrariness is met when torsion is allowed to propagate precisely because the very propagation of torsion requires conditions on the positivity of mass and energy which fix the sign of the torsion-spin potential in an inevitable way.

To see that indeed this leads to singularity avoidance, we notice that the kinetic, mass and potential terms scale according to $l^{4}$, $l^{3}$ and $l^{6}$ and so, for field densities that are large enough, as one would expect in the case of the singularity formation, there would be a threshold beyond which the above condition would approximate to
\begin{eqnarray}
&-\frac{X^{2}}{M^{2}}(|i\overline{\psi}\boldsymbol{\pi}\psi|^{2}\!+\!|\overline{\psi}\psi|^{2})
\!\geqslant\!0
\end{eqnarray}
and thus it would be violated: failing the hypotheses, the Penrose-Hawking singularity theorems would not engage.

On the other hand however, these arguments work for spinors satisfying either $i\overline{\psi}\boldsymbol{\pi}\psi\!\neq\!0$ or $\overline{\psi}\psi\!\neq\!0$ and not if
\begin{eqnarray}
&i\overline{\psi}\boldsymbol{\pi}\psi\!=\!\overline{\psi}\psi\!=\!0
\label{singular}
\end{eqnarray}
in which case gravitational singularities may form.

We recall that in general, spinor fields can be classified in terms of the Lounesto classification, in terms of which spinors possessing at least one of the two bi-linear scalars are called regular while those with $i\overline{\psi}\boldsymbol{\pi}\psi\!=\!\overline{\psi}\psi\!=\!0$ are the singular spinors \cite{daSilva:2012wp,Cavalcanti:2014wia}: the interesting fact we want to point out is that in Lounesto classification the character of being singular is due to constraints on the spinor and restrictions on its transformation properties, but it does not have anything to do with the gravitational tendency to form singularities as a consequence of gravitationally collapsing matter distributions into a region of vanishing volume in general. Nevertheless, as (\ref{singular}) clearly shows, a spinor that is singular in Lounesto classification is also a spinor that forms gravitational singularities after all.

The reasoning we have followed relies on the fact that we can always make the effective approximation, and that is that the densities beyond which the energy conditions fail are much smaller than the density beyond which the condition (\ref{a}) stops being valid, but of course it might happen that this is not the case, and that we cannot use the effective approximation in treating (\ref{conditionspecial}); were this the case, we would see no way to approach the problem other than finding some complete set of exact solutions.

Indeed, there are two instances in which exact solutions have been found, with no singularities \cite{Cianci:2015pba, Cianci:2016pvd}.

Whether this is a feature of all exact solutions we can not say, so in this case the problem is open.
\section{Conclusion}
In this paper, we have shown that for the most general renormalizable propagating torsion axial-vector massive field in gravity with spinors, the torsion-spin interactions are always attractive, and if the effective approximation remains valid at large enough densities, gravitational singularities are not bound to form; if they do not, then we know of no way to solve such a problem but finding exact solutions, and despite these being difficult to obtain, we have reported two instances in which exact solutions for a system of spinors in interactions with their own gravity field have actually been found, and both of them were in fact free of singularity: so, either because in the effective approximation singularities are impossible, or because in general the problem should be solved exactly and so far as we know singularities do not appear, in any case we may conclude that singularities do not constitute the problem that they are supposed to be. To summarize, the pivotal point of the argument is that the torsion-spin interactions are always attractive: attractive interactions may favour the clustering of spinors, but in doing so they do fill the space-time with negative potential energy, decreasing the energy source in the gravitational equations, determining the space-time curvature to relax, thus provoking a lesser tendency of the curvature to form singular points. Such an attractive interaction between spin and torsion has at its core the fact that the torsion tensor is in fact a Proca axial-vector massive field that can propagate.

As we have already discussed, the fact that the torsion is massive and propagating requires the torsion mass, in the torsion field equations, and the torsion energy density tensor, in the gravitational field equations, to be positive defined: with no propagating torsion, there is no sense in insisting on the positivity of energy and mass, so no sign can be fixed in the interaction. Also the fact that torsion is an axial-vector field is important, because in this case identity $-\overline{\psi}\boldsymbol{\gamma}^{\mu}\boldsymbol{\pi}\psi \overline{\psi}\boldsymbol{\gamma}_{\mu}\boldsymbol{\pi}\psi\!=\!
|i\overline{\psi}\boldsymbol{\pi}\psi|^{2}\!+\!|\overline{\psi}\psi|^{2}$ shows how a sign inversion occurs: instead, were torsion a vector field, identity $\overline{\psi}\boldsymbol{\gamma}^{\mu}\psi \overline{\psi}\boldsymbol{\gamma}_{\mu}\psi\!=\!
|i\overline{\psi}\boldsymbol{\pi}\psi|^{2}\!+\!|\overline{\psi}\psi|^{2}$ would give no sign inversion, and the force would be repulsive. Henceforth, all is due to the fact that torsion is a propagating massive axial-vector field, precisely as we have discussed here.

Opportunities for extensions can be sought in making more detailed studies on stability conditions: considering astrophysical systems, like the collapse of neutron stars, one may assume that all neutrons form condensates, and when this occurs, in (\ref{conditionveryspecial}) we may take the non-relativistic approximation, yielding $|2\overline{\psi}\psi|\!=\!mM^{2}\!/\!X^{2}$ as a condition for the equilibrium. Would this equilibrium be stable?

This condition resembles the degeneracy pressure that balances the gravitational collapse ensuring the stability of neutron stars, which in quantum mechanics is granted by Pauli exclusion. Could this mechanism have anything to do with the Pauli exclusion principle?

All present considerations are general and thus should they also apply for a black-hole?


\begin{thebibliography}{40}
\bibitem{C1} 
E.Cartan,
\textit{Annales Sci.Ecole Norm.Sup.} \textbf{40}, 325 (1923).
\bibitem{C2} 
E.Cartan,
\textit{Annales Sci.Ecole Norm.Sup.} \textbf{41}, 1 (1924).
\bibitem{C3} 
E.Cartan,
\textit{Annales Sci.Ecole Norm.Sup.} \textbf{42}, 17 (1925).
\bibitem{C4} 
E.Cartan,
\textit{Compt.Rend.Acad.Sci.} \textbf{174}, 593 (1922).
\bibitem{S} 
D.W.Sciama, in \textit{Recent Developments in\\
General Relativity} (Oxford, 1962).
\bibitem{K}
T.W.B.Kibble, 
\textit{J.Math.Phys.} \textbf{2}, 212 (1961).
\bibitem{h-h-k-n}
F.W.Hehl, P.Von Der Heyde, G.D.Kerlick,\\
J.M.Nester, \textit{Rev.Mod.Phys.} \textbf{48}, 393 (1976).
\bibitem{kop}
W. Kopczynski,
\textit{Phys.Lett.A}\textbf{39}, 219 (1972).
\bibitem{taf}
J. Tafel, 
\textit{Phys.Lett.A}\textbf{45}, 341 (1973).
\bibitem{tra}
A. Trautman, 
\textit{Nature Phys. Sci.}\textbf{242}, 7 (1973).
\bibitem{s-h}
J.M.Stewart, P.Hajicek,
\textit{Nature Phys. Sci.}\textbf{244}, 96 (1973).
\bibitem{ke}
G.D.Kerlick,
\textit{Phys.Rev.D}\textbf{12}, 3004 (1975).
\bibitem{Inomata:1976wi}
A.Inomata,
\textit{Phys.Rev.D}\textbf{18}, 3552 (1978).
\bibitem{Fabbri:2011kq}
L.Fabbri,
\textit{Gen.Rel.Grav.}\textbf{45}, 1285 (2013).
\bibitem{Magueijo:2012ug} 
J.Magueijo, T.G.Zlosnik, T.W.B.Kibble,\\
\textit{Phys.Rev.D}\textbf{87}, 063504 (2013).
\bibitem{Khriplovich:2013tqa}
I.B.Khriplovich, A.S.Rudenko,
\textit{JHEP}\textbf{1311}, 174 (2013).
\bibitem{Alexander:2014eva}
S.Alexander, C.Bambi, A.Marciano, L.Modesto,\\
\textit{Phys.Rev.D}\textbf{90}, 123510 (2014).
\bibitem{Buchbinder:1985ym}
I.L.Buchbinder, S.D.Odintsov, I.L.Shapiro,\\
\textit{Phys.Lett.B}\textbf{162}, 92 (1985).
\bibitem{Carroll:1994dq} 
S.M.Carroll, G.B.Field,
\textit{Phys.Rev.D}\textbf{50}, 3867 (1994).
\bibitem{Fabbri:2014dxa}
L.Fabbri,
\textit{Int.J.Geom.Meth.Mod.Phys.}\textbf{12},1550099(2015).
\bibitem{daSilva:2012wp}
J.M.Hoff da Silva, R.da Rocha,\\
\textit{Phys.Lett.B}\textbf{718}, 1519 (2013).
\bibitem{Cavalcanti:2014wia}
R.T.Cavalcanti,
\textit{Int.J.Mod.Phys.D}\textbf{23}, 1444002 (2014).
\bibitem{Cianci:2015pba}
R.Cianci, L.Fabbri, S.Vignolo,\\
\textit{Eur.Phys.J.C}\textbf{75}, 478 (2015).
\bibitem{Cianci:2016pvd} 
R.Cianci, L.Fabbri, S.Vignolo,\\
\textit{Eur.Phys.J.C}\textbf{76}, 595 (2016).
\end{thebibliography}
\end{document}